\documentclass[aps, reprint, superscriptaddress, amsmath, amssymb, bibnotes, nofootinbib]{revtex4-2}

\usepackage{graphicx}
\usepackage{dcolumn}
\usepackage{bm}
\usepackage{float}
\usepackage[normalem]{ulem}  
\usepackage{xcolor}
\usepackage{amsmath, amsfonts}
\usepackage{hyperref}
\hypersetup{colorlinks=true}
\hypersetup{pdfauthor={Name}}
\hypersetup{urlcolor=violet}
\usepackage{lineno}
\usepackage{mathtools}
\usepackage{amsthm}
\usepackage{tcolorbox}
\usepackage{cuted}
\usepackage[version=4]{mhchem}
\usepackage{tikz}
\makeatletter
\newtcolorbox{mybox}[3][]
{
  colframe = #2!25,
  colback  = #2!10,
  coltitle = #2!20!black,  
  title    = {#3},
  #1,
}

\newcommand{\orcid}[1]{\href{https://orcid.org/#1}{\includegraphics[width=10pt]{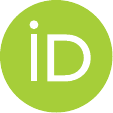}}}

\hypersetup{
  citecolor=blue 
}

\begin{document}

\title{New Signal of Atmospheric Tau Neutrino Appearance: \\ 
Sub-GeV Neutral-Current Interactions in JUNO}

\author{Stephan A. Meighen-Berger \orcid{0000-0001-6579-2000}\,}
\email{stephan.meighenberger@unimelb.edu.au}
\affiliation{School of Physics, The University of Melbourne, Victoria 3010, Australia}
\affiliation{Center for Cosmology and AstroParticle Physics (CCAPP), Ohio State University, Columbus, OH 43210, USA}

\author{John F. Beacom \orcid{0000-0002-0005-2631}\,}
\email{beacom.7@osu.edu}
\affiliation{Center for Cosmology and AstroParticle Physics (CCAPP), Ohio State University, Columbus, OH 43210, USA}
\affiliation{Department of Physics, Ohio State University, Columbus, Ohio 43210, USA}
\affiliation{Department of Astronomy, Ohio State University, Columbus, Ohio 43210, USA}

\author{Nicole F. Bell \orcid{0000-0002-5805-9828}\,}
\email{n.bell@unimelb.edu.au}
\affiliation{School of Physics, The University of Melbourne, Victoria 3010, Australia}
\affiliation{ARC Centre of Excellence for Dark Matter Particle Physics, School of Physics, The University of Melbourne, Victoria 3010, Australia}

\author{Matthew J. Dolan \orcid{0000-0003-3420-8718}\,}
\email{matthew.dolan@unimelb.edu.au}
\affiliation{School of Physics, The University of Melbourne, Victoria 3010, Australia}
\affiliation{ARC Centre of Excellence for Dark Matter Particle Physics, School of Physics, The University of Melbourne, Victoria 3010, Australia}

\date{3.~November 2023}

\begin{abstract}
We propose the first practical method to detect atmospheric tau neutrino appearance at sub-GeV energies, which would be an important test of $\nu_\mu \rightarrow \nu_\tau$ oscillations and of new-physics scenarios.  In the Jiangmen Underground Neutrino Observatory (JUNO; starts in 2024), active-flavor neutrinos eject neutrons from carbon via neutral-current quasielastic scattering.  This produces a two-part signal: the prompt part is caused by the scattering of the neutron in the scintillator, and the delayed part by its radiative capture.  Such events have been observed in KamLAND, but only in small numbers and were treated as a background.  With $\nu_\mu \rightarrow \nu_\tau$ oscillations, JUNO should measure a clean sample of 55 events/yr; with simple $\nu_\mu$ disappearance, this would instead be 41 events/yr, where the latter is determined from Super-Kamiokande charged-current measurements at similar neutrino energies.  Implementing this method will require precise laboratory measurements of neutrino-nucleus cross sections or other developments.  With those, JUNO will have $5\sigma$ sensitivity to tau-neutrino appearance in 5 years exposure, and likely sooner.
\end{abstract}

\maketitle


\section{Introduction}

Is the three-flavor neutrino mixing paradigm complete?  If not, this opens up the possibility of alternative explanations, which would be of profound importance for particle physics, astrophysics, and cosmology~\cite{Mohapatra:2005wg, Strumia:2006db, Mohapatra:2006gs, Iocco:2008va, Boyarsky:2009ix, Abazajian:2012ys, Bull:2015stt, Boyarsky:2018tvu, Dasgupta:2021ies}.  One key test is asking what disappearing active neutrinos transform into.  For solar neutrinos, for example, the long history of charged-current (CC) evidence for the disappearance of $\nu_e$~\cite{Cleveland:1998nv, GALLEX:1998kcz, SAGE:1999nng, SNO:2001kpb, Super-Kamiokande:2002ujc, Bellini:2011rx} was eventually met by the Sudbury Neutrino Observatory's neutral-current (NC) evidence for the appearance of a combination of $\nu_\mu$ and $\nu_\tau$~\cite{SNO:2002tuh, SNO:2011hxd}.  For atmospheric neutrinos, however, the long history of evidence for the disappearance of $\nu_\mu + \bar{\nu}_\mu$~\cite{Becker-Szendy:1992ory, Super-Kamiokande:2004orf, MINOS:2006foh, K2K:2006yov, T2K:2014ghj} has not yet been adequately met by evidence for the appearance of $\nu_\tau + \bar{\nu}_\tau$~\cite{Super-Kamiokande:2017edb, OPERA:2018nar, IceCube:2019dqi, IceCube:2020fpi}.  (Hereafter, when we say $\nu$, we mean $\nu + \bar{\nu}$, as they typically cannot be distinguished.)

Present results on $\nu_\tau$ appearance in GeV-range atmospheric-neutrino studies rely upon the facts that there is essentially no $\nu_\tau$ flux without oscillations, that an upgoing $\nu_\tau$ flux is generated through oscillations, and that tau leptons are produced in CC neutrino-nucleus interactions above 3.5~GeV~\cite{Jung:2001dh, Kajita:2010zz, Gaisser:2016uoy}.  While these $\nu_\tau$-induced events cannot be isolated individually, the fraction of such events can be measured statistically.  Data from Super-Kamiokande (Super-K) show that $\nu_\tau$ appearance is favored at $4.6 \sigma$~\cite{Super-Kamiokande:2017edb}; data from DeepCore, a dense infill detector of IceCube, support this at $3.2 \sigma$~\cite{IceCube:2019dqi}.  Separately, IceCube studies of near-PeV astrophysical neutrinos favor $\nu_\tau$ appearance at $2.8 \sigma$ based on events where there is enough time and/or distance separation between the events of tau-lepton creation and decay~\cite{IceCube:2020fpi}.  In combination, these results arguably exceed the usual $5\sigma$ criterion for discovery.  \textit{However, given the importance of fully testing the three-flavor paradigm, we need multiple results obtained under different physical conditions.}

In this paper, we introduce a new method, one that tests $\nu_\tau$ appearance via NC instead of CC interactions.  As with the NC-appearance technique used for solar neutrinos in the Sudbury Neutrino Observatory, this allows using neutrinos below the threshold for tau lepton production.  This method is made possible by a sensitive new experiment, the Jiangmen Underground Neutrino Observatory (JUNO), a scintillator detector with a fiducial volume of 18~kton that starts in 2024~\cite{JUNO:2015zny, JUNO:2021vlw, JUNO:2022mxj}.  For simplicity, when testing $\nu_\tau$ appearance, we take the null hypothesis to be simple $\nu_\mu$ disappearance, as done in the above-mentioned papers.  To test more specific scenarios, one would probe the fraction of $\nu_\tau$ appearance and would consider other constraints for, e.g., sterile-neutrino models.  We leave this for future work.

\usetikzlibrary{arrows.meta}
\begin{figure*}[t]
\begin{center}
{\tikz[remember picture]{\node(1AT){\includegraphics[width=0.44\textwidth]{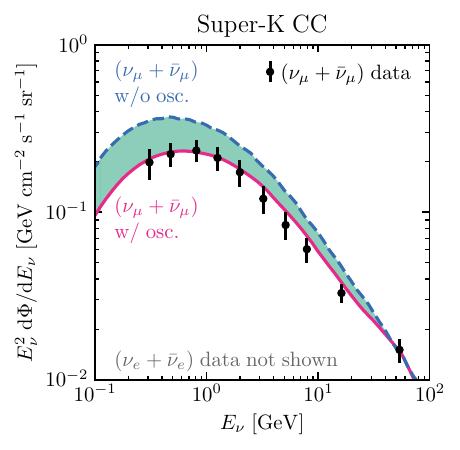}};}}
\tikz[baseline=-\baselineskip]\draw[-{Implies}, violet, line width=1pt, double, double distance=3pt] (-1,4) -- ++ (1,0);
{\tikz[remember picture]{\node(1AB){\includegraphics[width=0.44\textwidth]{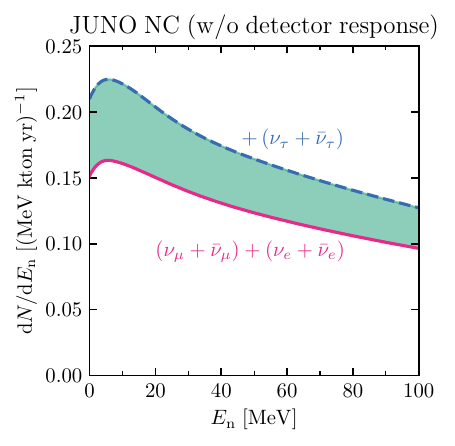}};}}
\end{center}
\caption{
\textbf{Left panel:} The measured $\nu_\mu$ spectrum at Super-K compared to our predictions without and with oscillations.  For clarity, we do not show the $\nu_e$ spectrum, which is hardly affected by oscillations; it is $\simeq$0.5 times as large as the $\nu_\mu$ spectrum without oscillations.  \textbf{Right panel:} The predicted neutron spectrum (without detector response; that is addressed in Sec.~\ref{sec:juno}) from NC events in JUNO under $\nu_\mu \rightarrow \nu_\tau$ versus $\nu_\mu$ disappearance.  Further details of the figure are explained below.
}
\label{fig:SKJUNO}
\end{figure*}

Figure~\ref{fig:SKJUNO} illustrates some of the essential ideas of our new method, which are as follows:
\begin{enumerate}

\item
All active flavors of atmospheric neutrinos induce NC interactions with nuclei.  In the sub-GeV neutrino energy range, which turns out to be the most important for our purposes, these interactions are quasielastic, often ejecting only a single neutron.  The NC interaction rate is relatively high.

\item
In JUNO, these neutrons induce a two-part signal.  The prompt part is caused by scattering of the neutron in the medium, primarily with protons.  The delayed part is caused by the neutron's eventual radiative capture, nearly always on a proton.  Both parts of the signal are detected with high efficiency.

\item
As the neutrino energies go down to 100~MeV, nearly all $\nu_\mu$ have oscillated, so for the $\nu_\mu \rightarrow \nu_\tau$ case, the flavor ratios become $\nu_e:\nu_\mu:\nu_\tau \simeq 1:1:1$.  For the case of simple $\nu_\mu$ disappearance, the NC signal rate in JUNO would then be 2/3 as large because $\nu_\tau$ would be absent.  For the larger neutrino energies we consider, where oscillations are less complete, this ratio is about 3/4.

\item We define expectations for JUNO's NC event rate without and with oscillations through Super-K's sub-GeV CC measurements of the $\nu_\mu$ and $\nu_e$ spectra at different arrival directions.  While this cancels flux uncertainties, precise laboratory measurements of neutrino-nucleus cross sections or other developments will be needed, as discussed in Sec.~\ref{sec:kamland}.

\end{enumerate}

Taking into account the details of neutrino oscillations, neutrino-nucleus interactions, and how events register in JUNO, we show that for $\nu_\mu \rightarrow \nu_\tau$ oscillations, JUNO should measure a clean sample of 55 events/yr in the detected energy range 11--29~MeV.  With $\nu_\mu$ disappearance, this would instead be 41 events/yr.  JUNO's statistical power will be increased if it can increase the detected energy range and exploit related NC channels.

In Sec.~\ref{sec:atmo}, we present our modeling of atmospheric neutrinos in Super-K, showing that we can reproduce their measured results well.  In Sec.~\ref{sec:kamland}, we do the same for KamLAND, a scintillator detector like JUNO but much smaller. KamLAND detected our proposed signal, but only in small numbers and treated as a background.  Having validated our modeling in these ways, in Sec.~\ref{sec:juno}, we present the details of our calculations for JUNO.  In Sec.\ref{sec:conclusion}, we conclude and discuss ways forward.


\section{Reproducing low-energy atmospheric data from Super-K}
\label{sec:atmo}

In this section, we review the fluxes and oscillations of atmospheric neutrinos, then model in detail their detectable signals in Super-K, which has the largest sample of well-reconstructed sub-GeV atmospheric neutrino events.  By validating our predictions against Super-K's energy and angular distributions, we establish a foundation for our predictions for KamLAND and JUNO.


\subsection{Atmospheric neutrino fluxes and oscillations}
\label{sec:atmo_subsec}

The low-energy atmospheric neutrino flux arises from the sequential decays of charged pions and muons produced in cosmic-ray interactions with nuclei in the upper atmosphere \cite{Engel:2011zzb, Gaisser:2016uoy}.  The flavor ratios before oscillations are thus   $\nu_e:\nu_\mu:\nu_\tau \simeq 1:2:0$.  For the fluxes, we use the site-dependent solar-cycle-averaged predictions of HKKM11~\cite{honda2011improvement} down to neutrino energies of 0.15~GeV, where they stop.  At lower energies, which barely matter for our results, we use similar results of Ref.~\cite{Zhuang:2021rsg} (which build on those of Ref.~\cite{Battistoni:2005pd}).

The primary flavor-change effect is due to $\nu_\mu \rightarrow \nu_\tau$ vacuum oscillations with the atmospheric parameters, $\sin^2 \theta_{23} = 0.55$ and $\Delta m^2_{32} = 2.44 \times 10^{-3} \; \mathrm{eV^2}$~\cite{Workman:2022ynf}.  The relevant oscillation length is
\begin{equation}
L_\mathrm{osc} (E_\nu) = \frac{4 \pi E_\nu}{\Delta m^2} \sim 10^3 \, \mathrm{km} \left( \frac{E_\nu}{\mathrm{GeV}} \right),
\label{eq:osclength}
\end{equation}
which should be compared (at order-of-magnitude level) to the production height in the atmosphere ($\sim10\,\mathrm{km}$), the distance to the horizon from Super-K ($\sim 10^3\,\mathrm{km}$) and the diameter of Earth ($\sim 10^4\,\mathrm{km}$).  When oscillations are fully developed and average out ($L/L_\mathrm{osc} \gg 1$), the flavor ratios are $\nu_e:\nu_\mu:\nu_\tau \simeq 1:1:1$.  Oscillations with the solar parameters happen only at larger distances and hence have little effect on the already equilibrated flavor ratios.

For neutrino flavor oscillations, we use {\tt nuCraft}, \cite{Wallraff:2014qka, nucraft}, which incorporates the dominant effects of vacuum oscillations and averaging over the atmospheric-neutrino production heights, as well as smaller effects due to matter-enhanced oscillations~\cite{Wolfenstein:1977ue, Mikheyev:1985zog, Akhmedov:1999ty, Krastev:1989ix}.

Figure~\ref{fig:SKJUNO} (left panel) compares our predicted $\nu_\mu + \bar{\nu}_\mu$ spectra without and with oscillations, showing that the ratio between them approaches a factor of two at low energies.  The spectrum shape follows from the proton spectrum and the kinematics of pion production near threshold~\cite{Gaisser:1991yk, Suliga:2023pve}.  Our predictions agree well with the angle-averaged neutrino spectra deduced by Super-K~\cite{Super-Kamiokande:2015qek}.  We caution that the Super-K points are not actual measurements, but rather follow from an inversion procedure that requires an ad-hoc regularization that produces large, correlated uncertainties.  Our predictions for the $\nu_e + \bar{\nu}_e$ spectra (not shown), which have only small changes due to oscillations, are also in good agreement with the Super-K results.


\subsection{Comparison to Super-K data}
\label{sec:spec_sk}

To further validate our oscillated flux model for Super-K, we use simulations to produce predictions that can be compared to their measured data in terms of directly measured energies~\cite{Super-Kamiokande:2017yvm}.  Super-K, a water Cherenkov detector with photomultiplier tubes on the walls, has a homogeneous fiducial volume of mass 22.5~kton, located in Japan at a depth of 1000 m (2700 m water-equivalent). For our purposes, Super-K's detection properties (energy and angular resolution, particle identification, and backgrounds) --- all of which we take into account in our calculations --- are so good that they cause only modest effects over the broad distributions in the data.  At low energies, the weak correlation between the lepton and neutrino directions (several tens of degrees) does have a significant effect on the angular distributions.  For the detected spectra as a function of  channel $k$, we use the following, which convolves three terms:
\begin{equation}\label{eq:rates}
    \frac{\mathrm{d}N_{\nu_{i,k}}}{\mathrm{d}E_\mathrm{det}} = 
    \frac{\mathrm{d}\phi_{\nu_i}}{\mathrm{d}E_\nu} \otimes
    \mathcal{C}_k(E_\nu, E_\mathrm{det})\otimes
    \epsilon_{i, k}(E_\mathrm{det}),
\end{equation}
where the first term is the oscillated neutrino spectrum for flavor $i$ (from the previous subsection), the second connects a neutrino energy to a range of detected energies, and the third is the detection efficiency (mostly due to analysis cuts as opposed to detector response).

To calculate the second term (detector response), we begin by simulating neutrino interactions in water (which are primarily with nuclei) with {\tt GENIE 3.2.0} with tune G18\_10a\_02\_11b, which is based on a local Fermi-gas model and an empirical meson-exchange model~\cite{Andreopoulos:2009rq, Andreopoulos:2015wxa, GENIE:2021zuu}.  (Figure~\ref{fig:genie_xsec} in the Appendix shows the most important total neutrino-oxygen cross sections.)  In addition to giving the interaction probabilities, {\tt GENIE3} also gives the full kinematic distributions of the final-state particles.  Most of the incoming neutrino energy is transferred to the outgoing charged or neutral lepton, which is mostly emitted in the forward direction, but the intrinsic energy and angular distributions are broad.  Next, we simulate the propagation of the final-state particles in Super-K using {\tt GEANT4}~\cite{GEANT4:2002zbu}.  This allows us to track the energy deposition of the primary particles as well as the creation and propagation of secondary particles.  For these combined simulations, we generate $10^7$ interactions, following an injection spectrum of $1/E$, which evenly samples in the log of energy.  We then reweight these events according to the atmospheric neutrino spectra.

\begin{figure}[t]
\centering
\includegraphics[width=0.97\columnwidth]{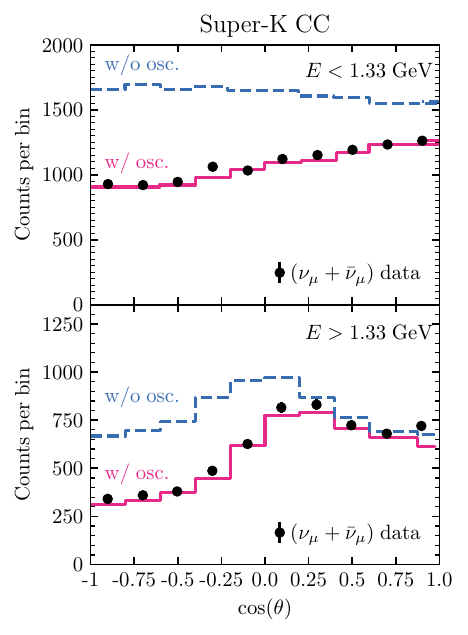}
\caption{
Zenith-angle distributions for sub-GeV (top panel) and multi-GeV (bottom panel) muon-neutrino events in Super-K (328 kton-yr exposure), compared to our predictions, showing very good agreement.  The statistical uncertainties are tiny, but there is an overall systematic uncertainty of $\sim$25\% (not shown). 
}
\label{fig:sk_mu_like}
\end{figure}

For the third term (efficiencies), we closely follow Ref.~\cite{Super-Kamiokande:2005mbp} to reproduce Super-K's analysis cuts and event classifications.  They divide sub-GeV and multi-GeV events at a visible-energy boundary of 1.33~GeV.  For fully contained events in the sub-GeV range, we consider muon decays with zero or one electron in the final state. Super-K's analysis cuts lead to identification efficiencies of 80\% for $\mu^+$ events and 63\% for $\mu^-$ events, where the difference is due to $\mu^-$ capture on nuclei, which leads to a lower efficiency because then the muon decay electron is not detected. The detection efficiency is 96\% (80\%) for both $\mu^+$ and $\mu^-$ for fully- (partially-) contained events. For high-energy events, we consider both fully- and partially-contained events, taking into account their spatial energy deposition and the detector geometry. In the later phases of Super-K, these efficiencies were improved. For example, between Phase-III and Phase-IV via new electronics \cite{Nishino:2009zu}, which improved the tagging efficiency of Michel electrons from 73\% to 88\%. This leads to our overall count prediction being slightly lower than the data.

Figure~\ref{fig:sk_mu_like} shows our predicted  zenith-angle distributions for muon-neutrino events.  As expected, the effects of neutrino oscillations are large, especially at low energies and long baselines ($\cos\theta_{\rm z} = -1$ corresponds to upgoing events).  The agreement of our predictions with data~\cite{Super-Kamiokande:2017yvm} is very good.  We find similar agreement for electron-neutrino events (not shown).  Together, this means that we have robust predictions for the spectra of sub-GeV atmospheric neutrinos without and with oscillations.  We note that the oscillation parameters have been independently and precisely determined by laboratory experiments, removing a degeneracy in interpreting the atmospheric-neutrino data.


\section{Reproducing low-energy atmospheric data from KamLAND}
\label{sec:kamland}

In this section, we focus on sub-GeV atmospheric NC interactions in scintillator detectors.  KamLAND has detected such events, but treated them as a background~\cite{KamLAND:2008dgz, Gando:2020cxo}.  To exploit them as a signal, detailed theoretical calculations are needed.  Reproducing the KamLAND data is a precondition to making accurate predictions for JUNO, which is much larger.


\subsection{NC interactions and signals}

For our predictions for KamLAND, we follow an approach similar to that of Sec.~\ref{sec:atmo} for Super-K, noting key differences below.  We take into account neutrino oscillations with {\tt nuCraft}, neutrino-nucleus interactions with {\tt GENIE3} (see the cross sections in Fig.~\ref{fig:genie_xsec} in the Appendix), and particle propagation with {\tt GEANT4}.  As above, we generate a large number of simulated interactions.

KamLAND is a liquid-scintillator experiment with a spherical active volume of 1~kton, located in Japan at a depth of 1000 m (2700 m water-equivalent)~\cite{Gando:2020cxo}.  The scintillator is composed of 80\% dodecane, 20\% pseudocumene, and 1.36 g/l PPO (2,5-diphenyloxazole) for fluorescence.  The nuclear mixture is 85\% C and 15\% H~\cite{Suzuki:2014woa}.  At the center of the active volume is a small balloon (radius 1.5 m before 2018, 1.9 m thereafter) with xenon-loaded liquid scintillator for double beta decay studies.  The fiducial volume for other studies is defined as a 5.5 m sphere around KamLAND's center, excluding certain regions around and above the small balloon.  Relative to water-Cherenkov detectors~\cite{Suekane:2004ny, Hyper-Kamiokande:2022smq}, liquid-scintillator detectors have a much larger detected photoelectron yield per MeV~\cite{Suzuki:2014woa, Gando:2020cxo}.  This improves energy measurements and makes it easy to detect neutron radiative captures, but the isotropic nature of the scintillation emission obscures event topologies and directions.  We take KamLAND's excellent energy and position resolution into account, though doing so has only modest effects.

In our simulations, we follow all possible final states, though we apply cuts as described below, after which the primary underlying interaction is NC quasielastic scattering of neutrons in carbon nuclei,
\begin{equation}
\nu + \ce{^{12}C} \rightarrow \nu + n + \ce{^{11}C^*},
\label{eq:nc_channel}
\end{equation}
which is the same for all neutrino flavors.  For $\bar{\nu}$, there is an indistinguishable NC interaction (the same for all antineutrino flavors), though with a smaller cross section (see Fig.~\ref{fig:genie_xsec} in the appendix) and somewhat different kinematics \cite{Meucci:2008zz}, compared to the neutrino case.  We always consider the sum $\nu + \bar{\nu}$.

Figure~\ref{fig:SKJUNO} (right panel) shows the initial spectrum of the neutrons in JUNO (similar for KamLAND).  A neutron is ejected with an initial kinetic energy of $\sim$$E_\nu^2/M_n$ (typically below a few hundred MeV), where we invoke non-relativistic kinematics and $M_n$ is the neutron mass.  The spectrum is falling primarily because of the cuts we apply and the nature of the differential cross section, which favors low neutron energies; the peak at a few MeV is due the falling atmospheric spectrum and cross section at low energies, plus nuclear effects.  This spectrum was also predicted in Ref.~\cite{JUNO:2022lpc, Cheng:2020aaw}, where it was considered only as a background for other searches in JUNO.  Our results are in reasonable agreement with theirs, though they use older simulations for the neutrino-nucleus interactions.  Compared to the energies relevant for us, they focus more on lower energies, where the neutrino-nucleus model differences are largest and where nucleon spectra due to nuclear de-excitations are more important.

Starting from our complete simulation results, we impose analysis cuts that match those used in KamLAND's experimental analyses~\cite{KamLAND:2011bnd, KamLAND:2021gvi}.  These criteria, plus selecting the energy range of interest for the prompt energy deposition to be 7.8--31.8~MeV, greatly reduce contributions from interactions besides those in Eq.~(\ref{eq:nc_channel}).  We select for two-part coincidence events with a prompt energy deposition and a delayed single neutron capture.  The parts of the events must be separated by less than $1000\;\mu$s in time (the mean is $\sim$210 $\mu$s) and 160~cm in space (the mean is $\sim$60~cm).  We also require that there are no other separable parts to the event, such as muon decays.  We do not attempt to identify nuclear final states through delayed decays.  We find that interactions different from Eq.~(\ref{eq:nc_channel}) contribute less than 10\% to the final event counts, as found in Refs.~\cite{KamLAND:2011bnd, KamLAND:2021gvi}.  We take into account KamLAND's livetime fraction of $\sim$80\% (due to spallation cuts following cosmic-ray muons) and their analysis efficiency of 73\% (due to requiring that both parts of the event be within the fiducial volume).

There are key differences in the underlying physics relative to Sec.~\ref{sec:atmo}, all of which we take into account.  A first difference is that here the prompt energy deposition is complicated compared to a single charged lepton with only continuous ionization losses.  A fast neutron undergoes many scatterings, including inelastic interactions that break apart carbon, as well as elastic interactions, where those with carbon primarily change the neutron's direction but not its energy and those with hydrogen do the opposite.  Of these processes, $n + p \rightarrow n + p$ is the most important for slowing the neutron, due to the equal masses.  Separately, the residual nucleus from the initial neutrino interaction (or those struck during neutron propagation) may be left in an MeV-range nuclear excited state that decays instantaneously, typically by gamma-ray emission, though sometimes with nucleon emission.  The gamma rays undergo Compton scattering or, less commonly, pair production.  All of this is included in the prompt energy deposition, but on average is only a small effect.

A second difference is that all of the prompt energy deposition is combined into isotropized and undifferentiated scintillation light.  Importantly, the light produced by heavy, nonrelativistic particles like hadrons with chsarge $Z$ and speed $\beta$ is reduced (``quenched") relative to that produced by relativistic electrons.  When the ionization energy loss rate, which is $\sim$(2 MeV/g/cm$^2$)~$Z^2/\beta^2$, is large, then collisional de-excitation of scintillator molecules becomes important relative to radiative de-excitation.  We account for quenching as follows~\cite{Birks:1951boa, Chou:1952jqv}:
\begin{equation}
E_\mathrm{equiv}= 
\int\limits_0^E
\frac{S \, \mathrm{d}E}
{1 + kB \left(\frac{\mathrm{d}E}{\mathrm{d}x}\right) + C \left(\frac{\mathrm{d}E}{\mathrm{d}x}\right)^2},
\label{eq:birk}
\end{equation}
which gives the electron-equivalent energy, $E_{\rm equiv}$, of the scintillation light produced by a single hadron of energy $E$.  Here ${\mathrm{d}E}/{\mathrm{d}x}$ is the energy loss rate, $S$ is the scintillation efficiency, and $kB$ and $C$ are free parameters. We use the values measured by KamLAND~\cite{Yoshida:2010zzd}: $kB = 7.79\times10^{-3}\;\mathrm{g/cm^2/MeV}$ and $C = 1.64\times 10^{-5}\;\mathrm{(g/cm^2/MeV)^2}$.  For protons with recoil energy 1, 10, and 100 MeV, the electron equivalent energies are 0.2, 7, and 89~MeV, respectively.  For the prompt energy deposition, we add the electron-equivalent energies of all hadrons produced by propagation of the final-state neutron; the contribution from protons is dominant.

A third difference is the eventual capture of the final-state neutron.  A fast neutron initially loses energy quickly; once it reaches thermal energies, it scatters elastically for a relatively long time until radiative capture occurs.  Typically, this is on a proton ($n + p \rightarrow d + \gamma$), releasing a 2.2-MeV gamma ray; rarely, it is on a carbon nucleus, releasing a 4.9-MeV gamma ray.  Following KamLAND, we require that these gamma rays are in 1.8--2.6~MeV or 4.4--5.6~MeV, respectively, taking into account the effects of energy resolution.  We emphasize that we require a single detected neutron capture, cutting events with extra neutrons either due to the initial production or due to final-state particle propagation.

\begin{figure}[t]
\centering
\includegraphics[width=0.97\columnwidth]{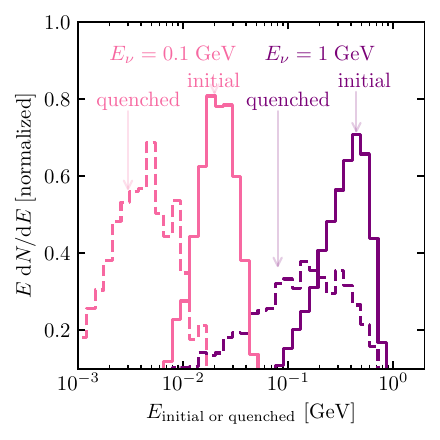}
\caption{
Neutron energy distributions --- initial and quenched total deposition --- for two example neutrino energies.
}
\label{fig:technical_stuff1}
\end{figure}

The backgrounds in our energy range of interest are low.  It may be possible to broaden this energy range beyond 7.8--31.8 MeV and thus increase the signal counts.  For the prompt energy deposition, the time profile of the scintillation light arriving at the photomultiplier tubes is different for hadron versus electron energy deposition. Pulse-shape discrimination techniques could thus help suppress the backgrounds, which dominantly have electrons.  Tagging the ground-state decay of $\ce{^{11}C}$ (which has a half-life of $20.4$ min and a beta-decay $Q$-value of 1.982~MeV~\cite{nudat3}) would cleanly isolate the interaction in Eq.~(\ref{eq:nc_channel}).  While this would be challenging, Borexino tagged such decays following cosmic-ray muon spallation~\cite{Galbiati:2004wx, Borexino:2006dsw}.  As an intermediate step, it should be possible to reject some events with other nuclear final states, due to their distinctive decays.  Last, it may also be possible to obtain crude directionality from the vector spatial separation between the $\ce{^{11}C}$ decay and the neutron capture, building on ideas in Refs.~\cite{Beacom:1998fj, CHOOZ:1999hgz, Vogel:1999zy}.  Novel reconstruction techniques may also help with directionality~\cite{Yang:2023rbg}.

Figure~\ref{fig:technical_stuff1} shows the energy distributions, without and with quenching, produced by neutrons of two example neutrino energies.  For the distributions without quenching, we show the initial neutron energy, which is very close to what will be deposited in the medium because losses due to neutrinos are minimal.  The spread of the distribution is due to the kinematics of the differential cross section, and is affected by the Fermi motion of the initial nucleons and by intranuclear scattering of final-state nucleons.  For the distributions with quenching, we show the equivalent electron energy, taking into account both the complicated scattering processes the neutron induces and the reduced scintillation efficiencies for hadrons.  The effects of quenching are nonlinear, being stronger for lower hadron (and hence neutrino) energies.


\subsection{Comparison to observed data}

As noted, KamLAND observed these atmospheric NC interactions in 7.8--31.8~MeV~\cite{KamLAND:2011bnd, KamLAND:2021gvi}, but treated them as a background in searches for low-energy $\bar{\nu}_e + p \rightarrow e^+ + n$ signals, e.g., from the diffuse supernova neutrino background.  Those CC events also have a two-part coincidence of a prompt energy deposition followed by a single neutron capture.  For such searches, atmospheric NC interactions are more relevant than atmospheric CC interactions because the former are more concentrated at low detection energies due to kinematics and quenching.  Outside KamLAND's energy range of interest, backgrounds due to spallation, reactor and atmospheric CC events are much larger.

\begin{figure}[t]
\centering
\includegraphics[width=0.97\columnwidth]{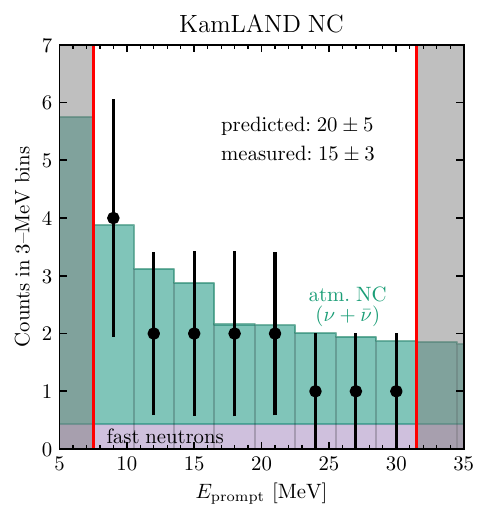}
\caption{
Spectrum of KamLAND's atmospheric NC events in 7.8--31.8~MeV (6.72~kton-yr exposure), compared to our predictions that take into account the full detector response.  For the KamLAND data, we have subtracted backgrounds due to spallation, reactor, and atmospheric CC events (all larger in the gray regions), plus rebinned the spectrum.
}
\label{fig:kamland_counts}
\end{figure}

Figure~\ref{fig:kamland_counts} shows our predicted atmospheric NC signal spectrum compared to 6.72~kton-yr of KamLAND data. We predict $17 \pm 4$ events in this energy range.  On top of this, we expect three fast neutron events due to muon interactions outside the active volume~\cite{KamLAND:2021gvi}, which we add to the sample.  Within this energy range, the tails of other backgrounds are small and well predicted, and we subtract their contributions.  We thus predict $20 \pm 5$ total events, while KamLAND observed  $15 \pm 3$~\cite{KamLAND:2021gvi}.  The agreement is very good, including for the shape, even without taking into account systematic uncertainties --- primarily on the fluxes and neutrino-nucleus cross sections --- which are expected to be a few tens of percent~\cite{Super-Kamiokande:2015qek, Cheng:2020oko, ZhouBeacom}.  This success further supports our modeling of low-energy atmospheric neutrinos.

\begin{figure}[b]
\centering
\includegraphics[width=0.97\columnwidth]{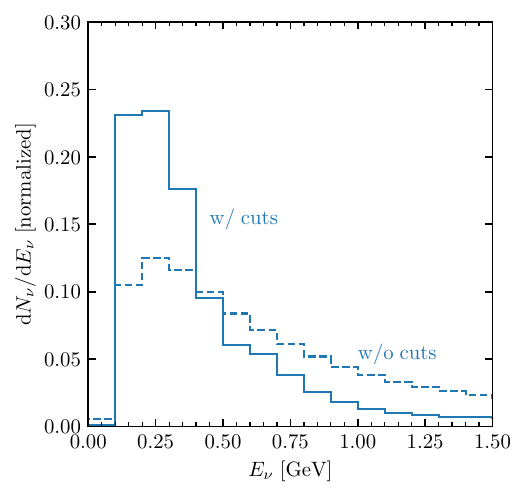}
\caption{
Distributions (separately normalized) of parent-neutrino energies for atmospheric NC events in KamLAND, without and with analysis cuts.  The sharpness of the step at low energies is an artifact of the binning.
}
\label{fig:technical_stuff2}
\end{figure}

Figure~\ref{fig:technical_stuff2} shows the distribution of parent neutrino energies for Eq.~(\ref{eq:nc_channel}), both for the total rate and for the rate after cuts.  To calculate the yield without cuts, we use
\begin{equation}
N_{th} =  \int \mathrm{d}E_\nu \, 
N_{t} \Delta t \frac{\mathrm{d}\Phi}{\mathrm{d}E_\nu}(E_\nu) \, \sigma_{NC}(E_\nu),
\label{eq:theory_estimate}
\end{equation}
where $N_{t}$ is the number of nucleons, $\Delta t$ is the livetime, we use the neutrino fluxes from Sec.~\ref{sec:atmo}, the NC cross sections shown in Fig.~\ref{fig:genie_xsec}.  For this ideal case, the parent-neutrino energy distribution is then determined by the integrand.  For the realistic case with cuts, we take into account all of the analysis cuts, including on the energy range.  This has a large impact on the shape of the parent-neutrino distribution, enhancing the low energy peak at $\sim$250~MeV and suppressing the contribution of neutrinos with energies above 500~MeV.  For these lower neutrino energies, the effects of neutrino oscillations are enhanced.

The relevant energies in Fig.~\ref{fig:technical_stuff2} are comparable to but not the same as those for the usual sub-GeV events in Super-K, which extend down to a visible energy of 250 MeV for muon neutrinos and 160 MeV for electron neutrinos~\cite{Super-Kamiokande:2015qek}.  For atmospheric neutrino data at lower energies, Super-K has only treated those events as a background~\cite{Super-Kamiokande:2002hei, Super-Kamiokande:2011lwo, Super-Kamiokande:2013ufi}, though Ref.~\cite{ZhouBeacom} finds good agreement with theoretical predictions.  It would be valuable for Super-K to develop detailed atmospheric-neutrino analyses down to the lowest energies.

\section{New predictions for tau-neutrino appearance in JUNO}
\label{sec:juno}

In this section, we present our calculations for JUNO and its sensitivity to atmospheric $\nu_\tau$ appearance.  With minor adjustments, our calculations closely follow those above for KamLAND, though JUNO is much larger.

JUNO's primary goal is high-precision measurements of reactor antineutrinos to determine the neutrino mass ordering, though it is a multipurpose detector~\cite{JUNO:2015sjr, JUNO:2015zny, JUNO:2021vlw, JUNO:2022lpc}.  The experiment, which is located in China at a depth of 700 m (1800 m water-equivalent), will start in 2024.  The active volume of 20~kton is a homogeneous sphere viewed by photomultiplier tubes, of which the fiducial volume is restricted to 18.3~kton to reduce fast-neutron and other backgrounds~\cite{JUNO:2022lpc}.  The scintillator is composed of linear alkyl-benzene, with 2.5 g/l PPO (2,5-diphenyloxazole) for fluorescence~\cite{JUNO:2020bcl}.  The nuclear mixture is 88\% C and 12\% H~\cite{JUNO:2015zny}. We hence adopt the same quenching parameter values as for our KamLAND calculation.

Beyond size, there are several relevant differences between JUNO and KamLAND.  Due to changes in the geomagnetic cut-off, we expect a $\sim$10\% smaller atmospheric neutrino flux at JUNO at the energies of interest~\cite{Zhuang:2021rsg}.  Due to the shallower depth, there is a higher flux of muons; the main concern is muon-induced fast neutrons from outside the active volume, but the huge size of JUNO allows effective shielding of those.  JUNO's yield of detected photoelectrons per MeV of energy deposited will be $\sim$4 times higher than for KamLAND~\cite{Suekane:2004ny,JUNO:2020bcl}.  And JUNO's better electronics will allow pulse shape discrimination techniques to separate signals and backgrounds. We restrict our analysis to prompt energies in the range 11--29~MeV to minimize backgrounds; we anticipate that detailed studies by JUNO will allow a broader energy range and thus a larger event rate.  To be conservative, we assume that the NC selection efficiency is 80\%, the same as the inverse beta decay selection efficiency~\cite{JUNO:2021vlw}. Realistically, this number should lie between 93\% and 99\%~\cite{KamLAND:2021gvi, Cheng:2020aaw}. Additionally, we assume a livetime efficiency of 80\% (the same as KamLAND), which can possibly be improved to 93\% \cite{JUNO:2021vlw}. Finally, we take JUNO's excellent energy and position resolution into account, though doing so has only modest effects.

\begin{figure}[t]
\centering
\includegraphics[width=0.97\columnwidth]{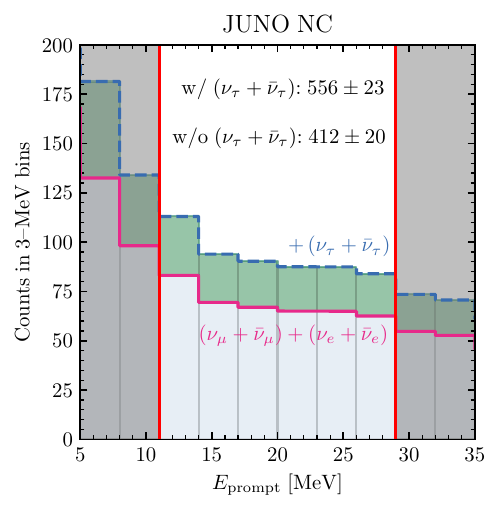}
\caption{
Predicted spectrum of atmospheric NC events in 11--29~MeV in JUNO (183~kton-yr exposure), including taking into account the full detector response.  We do not show the statistical uncertainties because they are evident and because our focus is on the integrated counts.
}
\label{fig:juno_counts}
\end{figure}

Figure~\ref{fig:juno_counts} shows our predicted energy spectrum for 10 years of JUNO data (183~kton-yr), where we have again selected events with a two-part coincidence of a prompt deposition followed by a neutron capture.  For the case of $\nu_\mu \rightarrow \nu_\tau$ oscillations, we expect 556 detected events with negligible backgrounds.  For the case of $\nu_\mu$ disappearance (the null hypothesis), we expect 412 events, which is smaller by 26\%.  Taking only statistical uncertainties into account in calculating the probability that 412 could fluctuate up to 556, this would give JUNO $7.1\sigma$ sensitivity to $\nu_\tau$ appearance in ten years (and $5\sigma$ sensitivity within five years).  With possible improvements to the analysis, these times would be shortened.

\begin{figure}[t]
\centering
\includegraphics[width=0.97\columnwidth]{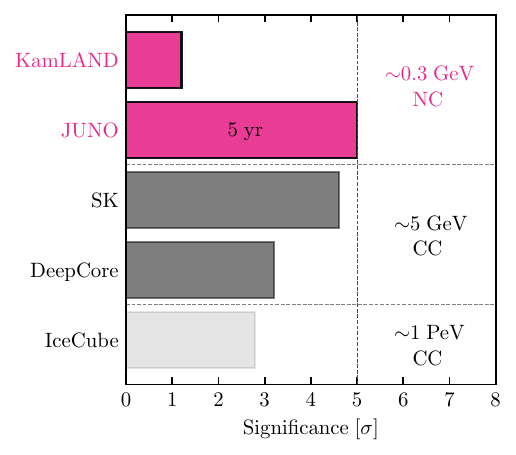}
\caption{
Our predicted sensitivities (statistical uncertainties only; see text) to $\nu_\tau$ appearance in atmospheric neutrino NC interactions (red bars) compared to present constraints based on atmospheric neutrino CC interactions (gray bars). We have also show current constraints on $\nu_\tau$ appearance from astrophysical measurements (light gray).
}
\label{fig:landscape}
\end{figure}

Figure~\ref{fig:landscape} compares our predicted sensitivities (ignoring systematic uncertainties) for the atmospheric NC signals in JUNO and KamLAND with existing results from experiments that rely upon tau lepton production from CC interactions. We see that JUNO will surpass the current SK sensitivity within 5 years; with an improved analysis this could occur sooner. Importantly, our new technique is complementary to existing approaches, probing NC interactions at much lower energies.

So far, we have ignored systematic uncertainties on the flux and cross sections, which are at the level of a few tens of percent~\cite{Super-Kamiokande:2015qek, Cheng:2020oko, ZhouBeacom}, as large as the difference we expect for $\nu_\mu \rightarrow \nu_\tau$ versus $\nu_\mu$ disappearance.  The flux uncertainty can be largely removed by basing the predictions for JUNO on Super-K data at comparable energies, as we have done.  Because Super-K has separately measured the atmospheric neutrino rates at all angles, the initial fluxes can be separated from the effects of oscillations, especially because the oscillation parameters are known from laboratory experiments.  However, there are significant cross section uncertainties because JUNO and Super-K have different compositions, plus one cross section is CC and the other is NC.  For simplicity, we discuss this in terms of the total cross sections, but it also applies to the differential cross sections.

Despite these difficulties with the cross section uncertainties, we are optimistic about ways forward.  First, we speculate that it may be possible to show that the CC neutrino-oxygen and NC neutrino-carbon uncertainties are largely correlated, in which case they would cancel in the comparison of JUNO and Super-K data. Additionally, JUNO could perform its own CC studies (which would require developing techniques for directionality), removing the dependence on Super-K data, so that the uncertainties would largely depend on comparing the CC versus NC neutrino-carbon cross sections, which are likely correlated.  Second, laboratory measurements of the cross sections could be made at accelerator near detectors, similar to measurements made by MiniBooNE~\cite{MiniBooNE:2010xqw} and T2K~\cite{T2K:2019zqh}.  A detailed uncertainty quantification based on existing data could prove more favorable than the few tens of percent we have assumed. Third, it may be possible to develop some crude directionality for the JUNO events, as noted above, so that comparison of upgoing and downgoing event counts would test $\nu_\tau$ appearance in these NC interactions.


\section{Conclusions and outlook}
\label{sec:conclusion}

A leading challenge in neutrino physics is to determine if the standard three-flavor paradigm is complete. A key test is to observe both the disappearance of active neutrinos due to flavor oscillations and the corresponding appearance of neutrinos of another flavor. A longstanding missing link is the observation of $\nu_\tau$ produced from oscillations of atmospheric $\nu_\mu$.
We have proposed the first practical way to test $\nu_\tau$ appearance at energies below the $\tau$ production threshold, using NC interactions. This method uses quasielastic scattering of neutrinos with carbon nuclei, with the ejection of a single neutron. These neutrons create a two-part coincidence signal in JUNO --- a prompt energy deposition from scattering of the neutron in the scintillator, followed by a delayed radiative capture of the neutron --- which greatly lowers backgrounds.  This signal has been observed in KamLAND (with low statistics) and predicted for JUNO, in both cases treated only as a background. For the first time, we have shown it to be a useful {\it signal}.  

The key obstacle to implementing our method is the neutrino-nucleus cross section uncertainties --- a problem that we believe will be surmountable in the near future.  Importantly, we expect that our method can be substantially improved. As discussed in Sec.~\ref{sec:kamland}, is it likely that the energy range can be expanded, increasing the statistics by a factor of a few. In JUNO, pulse-shape discrimination techniques and other advantages should allow decisive background rejection compared to KamLAND.  Also, JUNO should be able to use other NC interactions, for example, quasielastic NC interactions with protons~\cite{Chauhan:2021fzu}.  While this would not have a two-part coincidence signal, it should be possible to use pulse-shape discrimination to efficiently reject backgrounds.  If so, this would roughly match the statistics of our neutrino-neutron NC signal; it would also allow cross section uncertainties to be reduced through complementary measurements.  Such improvements would enable our new method to become a powerful technique to detect $\nu_\tau$ at sub-GeV energies.


\section*{Acknowledgements}

We are grateful for helpful discussions with Bhavesh Chauhan, Ivan Esteban, Yufeng Li, Kenny Ng, Sergio Palomares-Ruiz, Louis Strigari, and Bei Zhou.  This work was supported by the Australian Research Council through Discovery Project DP220101727.  It was also supported by the Research Computing Services and the Petascale Campus Initiative at The University of Melbourne.  J.F.B. was supported by National Science Foundation Grant No.\ PHY-2310018.


\appendix

\section{Neutrino cross sections}
\label{app:inter}

\begin{figure}[b]
\centering
\includegraphics[width=0.97\columnwidth]{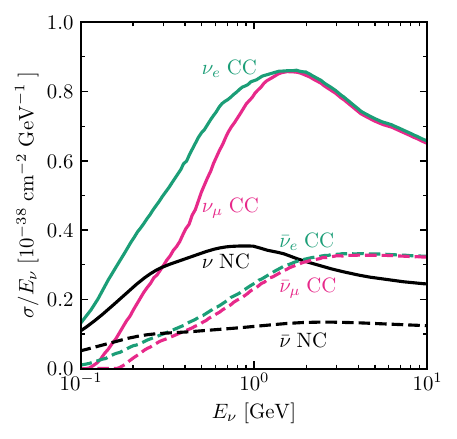}
\includegraphics[width=0.97\columnwidth]{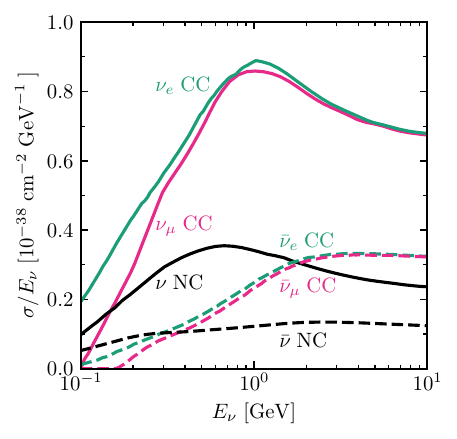}
\caption{
Key neutrino total cross sections as a function of neutrino energy, expressed as values per nucleon and per energy, as obtained from {\tt GENIE3}~\cite{GENIE:2021zuu}. \textbf{Top}: The oxygen cross sections. \textbf{Bottom}: The carbon cross sections.
}
\label{fig:genie_xsec}
\end{figure}

In this section, we show in Figure~\ref{fig:genie_xsec} the {\tt GENIE} results for the total cross sections (divided by neutrino energy) for both the neutrino-oxygen (Super-K) and neutrino-carbon (JUNO) cases.  Their similarity is encouraging from the perspective of potentially canceling their uncertainties in a ratio, but that is so far just a conjecture.

\clearpage
\bibliography{bibliography}

\end{document}